\documentclass[reprint,amsmath,amssymb,aps,prx,superscriptaddress,longbibliography]{revtex4-2} 
\usepackage{graphicx}
\usepackage{braket}
\usepackage{amsmath}
\usepackage{amsthm}
\usepackage{comment}
\usepackage{tikz}
\newtheorem{theorem}{Theorem}[section]

\newtheorem{definition}{Definition}
\newtheorem*{definition*}{Definition}

\usepackage[colorlinks,
bookmarksopen,
bookmarksnumbered,
citecolor=teal,
linkcolor=teal,
urlcolor=teal]{hyperref}
\usepackage{braket}

\usepackage{orcidlink}

\providecommand{\calC}{\ensuremath{\mathcal{C}}}

\providecommand{\calU}{\ensuremath{\mathcal{U}}}

\newcommand{\ie}{\textit{i.e.}\ }
\renewcommand{\qed}{\hfill \square}

\begin{document}

\title{Limits of Clifford Disentangling in Tensor Network States}

\author{Sergi Masot-Llima\orcidlink{0000-0002-6267-4911}}
\affiliation{Barcelona Supercomputing Center, Barcelona 08034, Spain}
\affiliation{Universitat de Barcelona, Barcelona 08007, Spain}

\author{Piotr Sierant \orcidlink{0000-0001-9219-7274}}
\affiliation{Barcelona Supercomputing Center, Barcelona 08034, Spain}

\author{Paolo Stornati~\orcidlink{0000-0003-4708-9340}}
\affiliation{Barcelona Supercomputing Center, Barcelona 08034, Spain}

\author{Artur Garcia-Saez  \orcidlink{0000-0003-3561-0223}}
\affiliation{Barcelona Supercomputing Center, Barcelona 08034, Spain}
\affiliation{Qilimanjaro Quantum Tech, 08019 Barcelona, Spain}

\date{ \today }

\begin{abstract}
Tensor network methods leverage the limited entanglement of quantum states to efficiently simulate many-body systems. Alternatively, Clifford circuits provide a framework for handling highly entangled stabilizer states, which have low magic and are thus also classically tractable. Clifford tensor networks combine the benefits of both approaches, exploiting Clifford circuits to reduce the classical complexity of the tensor network description of states, with promising effects on simulation approaches. We study the disentangling power of Clifford transformations acting on tensor networks, with a particular emphasis on entanglement cooling strategies. We identify regimes where exact or heuristic Clifford disentanglers are effective, explain the link between the two approaches, and characterize their breakdown as non-Clifford resources accumulate. Additionally, we prove that, beyond stabilizer settings, no Clifford operation can universally disentangle even a single qubit from an arbitrary non-Clifford rotation. Our results clarify both the capabilities and fundamental limitations of Clifford-based simulation methods.
\end{abstract}
\maketitle



\section{Introduction}
Quantum simulators~\cite{Lewenstein07, Bloch08,Cirac12, Altman21} and quantum computers~\cite{DiVincenzo00, Montanaro16, Cerezo21, Bharti22} offer controllable platforms for realizing quantum many-body systems and preparing states that, at sufficiently large qubit number, evade classical tractability, sharpening the prospect of quantum computational advantage~\cite{Preskill12frontier, Daley22practical}. 
Consequently, it is essential to advance classical descriptions of such states, which can be achieved by leveraging the characteristic structure~\cite{Waintal24} of physical many-body wavefunctions. 
At the heart of this structure lies quantum entanglement~\cite{Horodecki09quantum, Amico08entanglement}, the hallmark of correlations beyond those possible with classical systems. 
Whereas typical (Haar-random) states exhibit extensive, volume-law entanglement~\cite{Page93ent}, many physically relevant regimes, including ground states of gapped local Hamiltonians~\cite{Hastings07, Eisert10area, Laflorencie16} and certain non-equilibrium phases, display constrained entanglement, and are governed by area laws or slow (e.g., logarithmic) growth~\cite{Abanin19, Serbyn21, Sierant25rev}. 
Tensor Network (TN) methods~\cite{Vidal03a, Verstraete04, Verstraete08, Orus14, Haegeman16, Orus19, ran2020tensor, Banuls23} explicitly harness this limited entanglement to efficiently compress the quantum state, permitting a compact description of the wavefunction with complexity scaling with its entanglement content rather than the qubit number.

However, entanglement is not the sole determinant of classical simulation complexity. 
A prominent counterexample is found in stabilizer states generated by Clifford circuits; these states can exhibit extensive volume-law entanglement~\cite{Smith06typical, Dahlsten07distr} yet remain efficiently simulable via the Gottesman-Knill theorem~\cite{Gottesman98, Aaronson04improved}. 
This dichotomy motivates the framework of quantum resource theories~\cite{Chitambar19}, specifically the resource theory of non-stabilizerness (or quantum magic)~\cite{Veitch2014theresourcetheory, Heimendahl22, Haug23monotones}, which quantifies the deviation of a given quantum state from the manifold of stabilizer states. 
The field has been invigorated by the introduction of efficiently computable magic measures, most notably the Stabilizer Rényi Entropies~\cite{leone2022stabilizerrenyientropy, leone2024stabilizer}. 
Recent algorithmic breakthroughs in evaluating these measures for both Tensor Network states~\cite{haug2023quantifying, lami2023perfect, tarabunga23gauge, Tarabunga24mps, liu2025translational, tarabunga2023critical} and exact statevectors~\cite{Sierant26computing, Xiao26exp, Huang26fast, Tarabunga25} have opened a new window into many-body physics, providing fresh perspectives on equilibrium phases~\cite{tarabunga23gauge, Odavic23, Falcao25, Hoshino25sre, Timsina25rob} and non-equilibrium dynamics~\cite{turkeshi2024magic, Turkeshi25spectrum, Tirrito25anti, Falcao25non, Aditya25Mpemba, Aditya25spreading}.
One constructive approach to generate states with controlled amount of nonstabilizerness~\cite{Magni25complexity, Magni25anti, Magni25doped} is by forming superpositions of stabilizer states~\cite{Bravyi15, Bravyi16, Bravyi19}.

A recent approach to using complementarity between entanglement complexity and non-stabilizerness stems from asking whether the computational efficiency of Clifford circuits can be harnessed to extend the reach of Tensor Networks. 
Specifically, is it possible to utilize Clifford operations to ``disentangle'' a quantum state, yielding a residual wavefunction with sufficiently low entanglement to be efficiently compressed? 
This line of inquiry has driven the development of Clifford Tensor Network (CTN) techniques~\cite{stabTN, Nakhl25stab, Mello24MPO}. 
These hybrid architectures~\cite{Frau_2024, cemps_lami} aim to integrate the strengths of Tensor Networks, which excel at parameterizing local correlations, with the ability of Clifford circuits to represent the extensive volume-law entanglement characteristic of typical many-body states.
This paradigm has proven particularly effective for variational methods: it has been applied to ground state searches by augmenting the Density Matrix Renormalization Group (DMRG)~\cite{White93dmrg} with Clifford layers~\cite{Qian_2024}, and extended to non-equilibrium dynamics~\cite{Mello25dressed} via the time-dependent variational principle~\cite{Haegeman16}.
Central to these approaches is the mechanism of disentangling (or entanglement cooling)~\cite{Fan25disent, Fux25disent, Frau25conformal, Liu25cliff}, where the Clifford circuit effectively strips away long-range correlations, mapping the complex physical state into a low-entanglement form manageable by the tensor network backend.

In this work, we employ a paradigmatic model of doped Clifford circuits~\cite{Leone24doped} to probe the fundamental limits of entanglement cooling in beyond-Clifford states. Focusing on one-dimensional systems, where the CTN framework reduces to Clifford-augmented Matrix Product States (CAMPS)~\cite{cemps_lami, Fux25disent}, we provide critical insights into the capabilities and boundaries of these simulation tools. 

The remainder of this manuscript is organized as follows. Section~\ref{sec:cetn} establishes the theoretical background for CTNs and outlines the entanglement cooling protocols employed. In Sec.~\ref{sec:limitations}, we investigate the rigorous limitations of these protocols, proving that perfect Clifford disentangling is achievable only if the state, prior to the non-Clifford gate application, factorizes into a product state containing a Clifford subsystem. Section~\ref{sec:ent_exp} analyzes the entanglement accumulation in CAMPS under different classes of non-Clifford gates and examines the impact of bond dimension truncation on simulation fidelity. Finally, Sec.~\ref{sec:conc} offers concluding remarks and an outlook on future directions.


\section{Background}


\subsection{Clifford-based simulation methods}\label{sec:cetn}

\begin{figure*}[!t]
	\centering
	\includegraphics[width=0.85\linewidth]{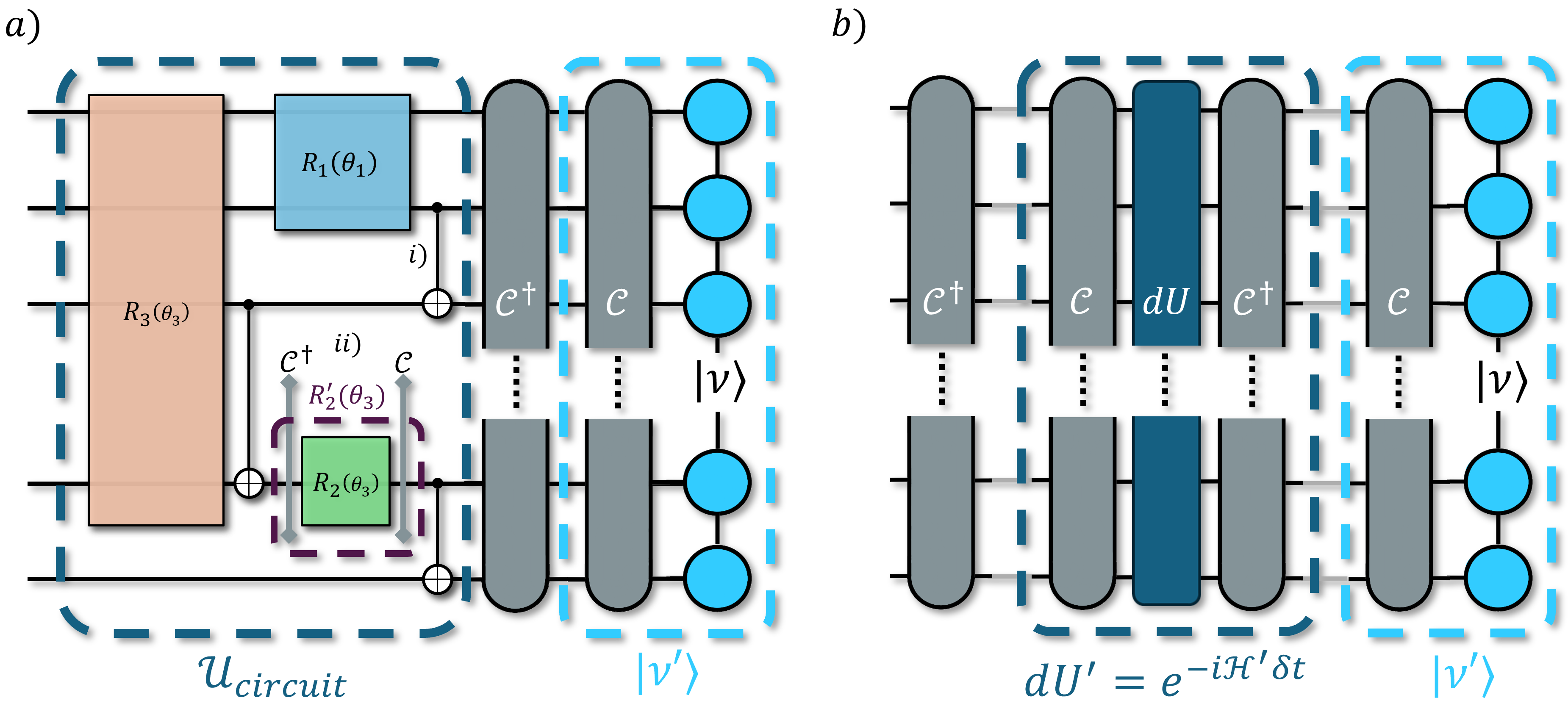}
	\caption{Simulations 
    with Clifford enhanced TN (MPS in the example). In a), we can find a Clifford unitary $\calC$ that reduces bond dimension $\chi$ of an MPS $\ket{\nu}$, yielding $\ket{\nu'}$. Any Clifford gate in the circuit, such as those marked i), can be absorbed into $\calC$ to obtain $\calC'$, whereas non-Clifford rotations must be conjugated with $\calC$ before being contracted to $\ket{\nu'}$. The search for a better $\calC$ can be repeated at any step of the circuit simulation. In b), the Hamiltonian evolution of a quantum state with MPS and time steps $dU$ can similarly be improved with a disentangling Clifford gate that decreases $\chi$ of $\ket{\nu}$, whereas the conjugate is contracted with the Hamiltonian, giving $dU'$.}
	\label{fig:ham_ev_cmps}
\end{figure*}

The representability of quantum states via a Tensor Network (TN) $\mathcal{T}$ is fundamentally constrained by the bond dimension $\chi$, which limits the entanglement capacity of the ansatz. This expressivity can be significantly enhanced by augmenting the TN with a Clifford transformation $\mathcal{C}$. Although Clifford circuits can generate extensive volume-law entanglement, they remain classically efficient to simulate. 
The resulting composite ansatz $(\mathcal{C},\mathcal{T})$ provides, instead, a versatile simulation framework. A universal circuit simulation with this structure~\cite{stabTN} processes Clifford gates by updating $\mathcal{C}$ and non-Clifford gates by updating $\mathcal{T}$. 
In the context of Hamiltonian dynamics~\cite{Mello25dressed}, this approach enables access to longer timescales at fixed $\chi$. 
This is achieved by identifying a Clifford unitary $\mathcal{C}$ that ``disentangles'' the state, thereby minimizing the bond dimension required for $\mathcal{T}$. 
Such procedure effectively shifts the complexity to the operator frame, transforming the Hamiltonian $\mathcal{H}$ into $\mathcal{H}' = \mathcal{C}^\dagger \mathcal{H} \mathcal{C}$. 
Crucially, since Clifford operations map Pauli strings to Pauli strings, $\mathcal{H}'$ retains the same number of terms and coefficients as the original Hamiltonian, serving as the effective generator for the subsequent time step (see Fig.~\ref{fig:ham_ev_cmps}).

From now on, we refer to these simulation methods as Clifford Tensor Networks (CTN) methods: 
\begin{definition}[Clifford Tensor Networks]\label{def:ctn}
	Clifford Tensor Networks are the family of quantum state simulation methods based on updating a Clifford transformation $\mathcal{C}$ and a tensor network $\mathcal{T}$ simultaneously, so that the simulated state $\ket{\psi}$ can contain more entanglement than that permitted with the bond dimension $\chi$ of $\mathcal{T}$. Denoting by $ \ket{\psi_\mathcal{T}}$, the quantum state of the system is represented as:
	\begin{equation}\label{eq:ctn}
		\ket{\psi} = \mathcal{C} \ket{\psi_\mathcal{T}}.
	\end{equation}
\end{definition}
The study of topological order and other features of many-body systems is also related to Clifford transformations, suggesting possible benefits from connecting the two areas~\cite{Fan25disent,hoshino2025stabilizerrenyientropyencodes,PhysRevA.110.062427}. Furthermore, the previous work in stabilizer Tensor Networks \cite{stabTN} has inspired directly studies on different circuit architectures like state injection~\cite{Nakhl25stab}, and there is also interest in understanding the role of stabilizer bases in general~\cite{Sun2025stabilizerground}. More traditional TN methods such as DMRG can and have been used on CTN~\cite{Qian_2024} 

To fully leverage these approaches, it is essential to integrate an entanglement cooling subroutine. Performed periodically during the simulation, this step optimizes $\mathcal{C}$ to compress the correlations in $\mathcal{T}$ without altering the underlying physical state, as detailed in the following section. Furthermore, our understanding of these methods benefits from parallel developments in magic-based simulation and surrogation techniques~\cite{p7xt-s9nz, dowling2025bridgingentanglementmagicresources, Mello24MPO}. These related approaches operate primarily in the Heisenberg picture, focusing on the efficient evolution of operators rather than states.


\subsection{Entanglement cooling in Clifford enhanced Tensor Networks}\label{sec:ent_cool}

The efficacy of CTN lies in their ability to offload entanglement storage from the tensor network to the Clifford operations. However, the partitioning of correlations between these two manifolds is not unique, and its optimal management is non-trivial. For instance, when encountering a Clifford gate in a dynamical simulation, one must decide whether to contract a gate into the tensor network or absorb it into the Clifford operation. To navigate this ambiguity, we employ an entanglement cooling strategy~\cite{Chamon14, True2022transitionsin}, which exploits the gauge freedom of the CTN. This process inserts a resolution of the identity, $\mathbb{I} = \mathcal{U}_C \mathcal{U}_C^\dagger$, into the state representation. The unitary $\mathcal{U}_C$ is applied to the tensor network to reduce its bond dimension $\chi$, while the inverse $\mathcal{U}_C^\dagger$ is absorbed into the Clifford frame. Since these operations cancel globally, the physical state remains invariant, yet the representational complexity can be lowered.

Implementing this protocol requires solving a complex optimization problem. Identifying the globally optimal disentangler $\mathcal{U}_C$ is computationally intractable due to the superexponential scaling of the Clifford group size with system size~\cite{Bravyi_2022}. Consequently, practical algorithms rely on local heuristics. A robust approach involves minimizing the entanglement entropy (specifically, the maximum entropy across all partitions) through a greedy search over the space of geometrically $k$-local Clifford gates—transformations acting on at most $k$ adjacent sites. The algorithm proceeds by performing sweeps up to depth $d$ across the system. This local optimization strategy was originally proposed in the context of MPS~\cite{cemps_lami} and utilized for Hamiltonian dynamics~\cite{Mello25dressed}, though its specific application to quantum circuit simulation was only recently formalized~\cite{Fux25disent}.

\begin{definition}[$K$-local entanglement cooling]\label{def:heur_EC}
	For an MPS-based CTN with ($\mathcal{C},\mathcal{T}$), the $k$-local heuristic entanglement cooling reduces the entanglement of $\mathcal{T}$ and thus its precision for a given $\chi$. It proceeds by iteratively sweeping over the sites of the MPS in groups of $k$-sites, contracting them with each Clifford transformation $C\in\mathcal{C}_k$ and evaluating the resulting entanglement entropy $S$:
    \begin{equation}\label{eq:vnentropy}
        S(\rho) = - \sum_i \rho_i \log_2 \rho_i,
    \end{equation} with $\rho_i$ the values of the singular value decomposition of a cut within those sites. The $C$ that shows better improvement is chosen and $\mathcal{C}$ updated with it, before moving to the next group; up to a depth of $d$ sweeps from side to side.
\end{definition}

\begin{figure}[b]
	\centering
	\includegraphics[width=\linewidth]{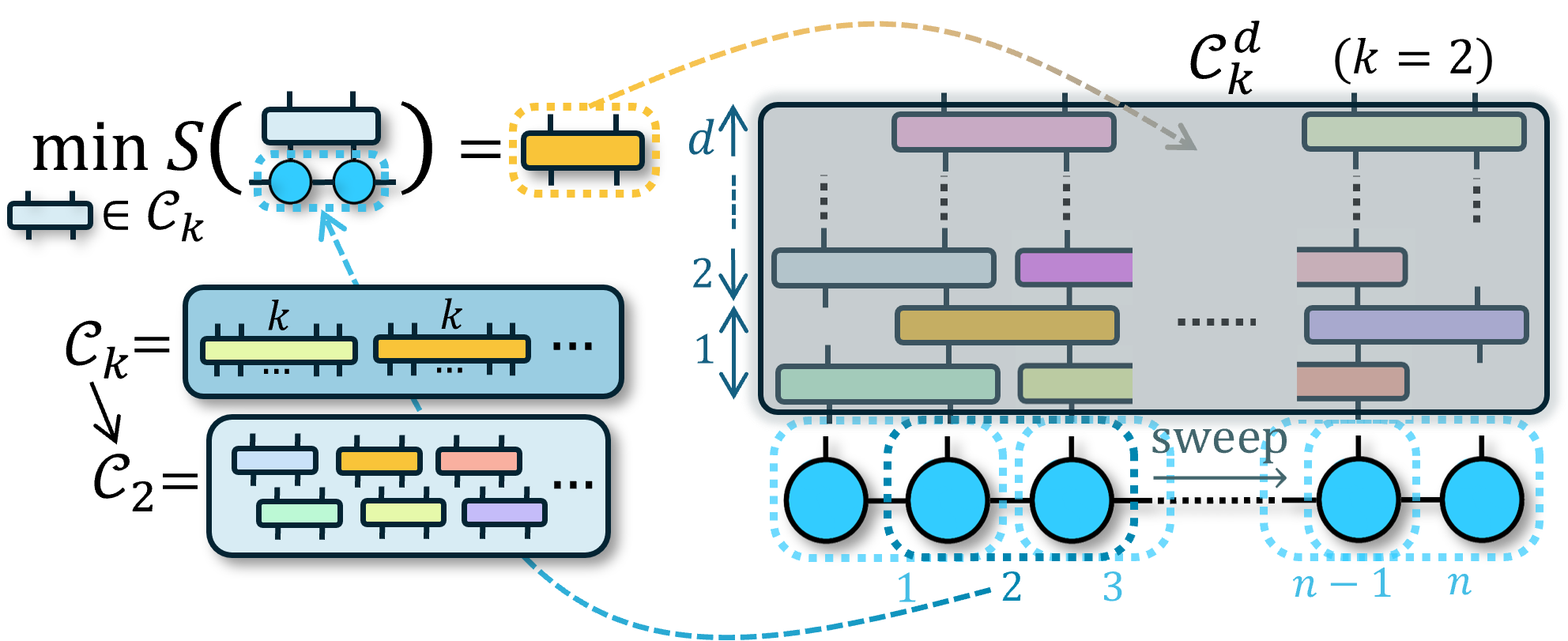}
	\caption{Heuristic $K$-local entanglement cooling illustrated for $k=2$. Proceeding in a sweep over the MPS, each pair (group of $k$) of sites are contracted with the elements of $\mathcal{C}_2$ ($\mathcal{C}_k$), to find the one with minimum entanglement entropy $S$, which is chosen for that step. The procedure stops after $d$ sweeps, what we call depth, and returns the resulting Clifford gate $C^d_2$ ($C^d_k$).}
	\label{fig:ent_cool_heur}
\end{figure}

The algorithmic procedure is schematically illustrated in Fig.~\ref{fig:ent_cool_heur}. While the protocol was originally formulated with $2$-local Clifford gates, extending it to $k > 2$ requires a precise definition of the optimization objective, specifically, whether to minimize the average entropy across the partition or to target the maximum entropy cut. To constrain the computational cost, the optimization is truncated after a fixed depth $d$ of sweeps. 
Furthermore, the search space can be dramatically pruned by exploiting the local unitary invariance of entanglement measures. Although the two-qubit Clifford group $\mathcal{C}_2$ contains $11520$ elements~\cite{kubischta2024}, this count is inflated by single-site Clifford transformations. Specifically, we define two Clifford unitaries $U$ and $V$ as equivalent ($U \sim V$) if they are related by local operations: $V = (L_1 \otimes L_2) U (R_1 \otimes R_2)$, where $L_i, R_i$ are single-qubit Clifford gates. Dividing out these local single-site contributions, which do not alter the entanglement, reduces the group to a set of distinct equivalence classes. By selecting one representative from each entangling class, the search space collapses to merely $20$ gates.

\begin{figure}[!ht]
	\centering
	\includegraphics[width=\linewidth]{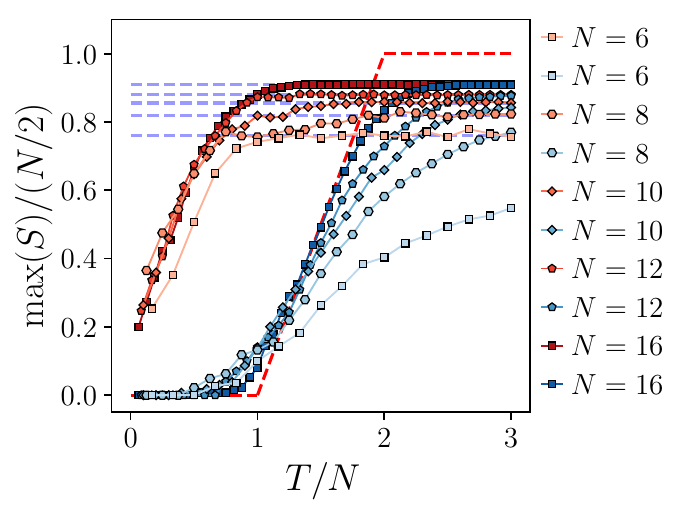}
	\caption{Progression of the maximal entanglement $max(S)$ across all sites of an sTN during the simulation of a Clifford+$T$-gates circuit, for different sizes $N$, and up to $3N$ $T$-gates. We compare circuits where the entanglement cooling in Fig.~\ref{fig:ent_cool_heur} was used (in blue) to those where it was not (red), and we normalize both axes with $N/2$ to compare different sizes better. Dashed blue lines correspond to the entropy bound $S_b(N)$~\cite{Page93entropy} for finite $N$, in ascending vertical order.}
	\label{fig:ent_cool_example}
\end{figure}

Figure~\ref{fig:ent_cool_example} demonstrates the efficacy of this heuristic using the sTN library~\cite{stabTN_code} on quantum circuits considered in~\cite{Fux25disent}, that is, comprising global random Clifford gates interspersed with $T$-gates that act as sources of non-stabilizerness in the system. In this benchmark, entanglement cooling is triggered after every non-Clifford operation to minimize the bond dimension $\chi$. While the frequency of cooling is a tunable parameter, evidence from Hamiltonian evolution studies (see Fig.~\ref{fig:ham_ev_cmps}) suggests that applying the protocol at every time step yields superior precision compared to intermittent application. A deeper theoretical understanding of the algorithm's convergence is needed to identify regimes where sparse cooling suffices.

Three distinct behaviors emerge in the disentangled entanglement entropy as a function of the $T$-gate density. These regimes correspond to the intervals $0 < T < N$, $N < T < 2N$, and $T \geq 2N$. Notably, in the first regime, the cooling procedure is highly effective, often reducing the entanglement to zero. In this limit, the heuristic optimization procedure converges to the \textit{exact} entanglement cooling protocols previously reported in~\cite{Fux25disent,exact_lami,Liu25cliff} and proved analytically in \cite{Fux25disent}.
\begin{definition}[Exact entanglement cooling]\label{prop:exact_EC}
For a CTN state $(\mathcal{C}, \mathcal{T})$ representing $n$ qubits, in which the tensor network component $\mathcal{T}$ factorizes such that qubit $i$ is in a single-qubit stabilizer state $|\sigma\rangle_i$ (an eigenstate of a Pauli operator $P_c$), let $U = \alpha \mathbb{I} + \beta P$ be a rotation generated by the Pauli string $P = \bigotimes_{k} P_k$. If the local Pauli operator $P_i$ acts non-trivially on the state (i.e., $P_i |\sigma\rangle_i \neq |\sigma\rangle_i$), then the global operation $U$ can be perfectly disentangled. Specifically, $U$ decomposes into a local rotation $\alpha \mathbb{I} + \beta P_i$ acting solely on qubit $i$, and a cascade of controlled-Pauli gates absorbed into the Clifford operation $\mathcal{C}$. These entangling gates are controlled by qubit $i$ in the $P_c$ basis, applying the corresponding Pauli $P_j$ to each target qubit $j \neq i$.
\end{definition}
We designate this protocol as exact because it derives from a deterministic, constructive decomposition rather than a variational optimization. Its structure exploits the conjugation of Pauli rotations by Clifford unitaries, as illustrated in Fig.~\ref{fig:ham_ev_cmps}(a)(ii) and further detailed in Appendix~\ref{app:exact_dis}. The convergence of the heuristic approach to this exact solution is non-trivial, given the discrete and high-dimensional nature of the Clifford group. In Sec.~\ref{sec:exact_equals_heur}, we demonstrate that this agreement is a natural consequence of the specific gate structure inherent to these algorithms. Conversely, in the second regime ($N < T < 2N$), entanglement accumulates at a rate comparable to the uncooled dynamics (shown in red in Fig.~\ref{fig:ent_cool_example}), a phenomenon we analyze in Sec.~\ref{sec:ent_exp}.

Finally, the entropy stabilizes in the third and last regime ($T \geq 2N$). There, we see a gap between the asymptotic value and the theoretical maximum entropy $S=1$, which corresponds to the approximate bound $S_b(N)=1-1/(N\log(2))$ on entropy for Haar-random states (that is, uniformly drawn~\cite{Mele2024introductiontohaar}) with finite $N$, as proven in Ref.~\cite{Page93entropy}. This upper limit, which we depict in dashed blue lines in Fig.~\ref{fig:ent_cool_example} for each $N$, increases monotonically with $N$ up to $S=1$ in the $N\to \infty$ limit.
In general, the entropy of these circuits reaches the value of Haar-random states soon after $N$ T-gates on average, which is delayed to soon after $2N$ T-gates when applying disentangling. It also reveals that the $2$-local heuristic cooling algorithm fails to find relevant improvements when the states are sufficiently close to the Haar-random limit.


\section{Limitations of CTNs}\label{sec:limitations}

There are multiple facets of entanglement cooling algorithms that warrant critical examination to establish their range of validity. In this section, we investigate the limitations of these protocols. Specifically, we rule out the feasibility of several conjectured improvements to both heuristic and exact disentangling strategies, thereby clarifying the fundamental constraints of CTNs.

\subsection{Heuristic disentangling with higher locality Clifford gates}

Once the structural simplicity of the state breaks down beyond the $N < T$ regime, the standard $k$-local heuristic yields only marginal returns, failing to arrest the asymptotic growth of entanglement. However, since the limitations of entanglement cooling are intrinsically tied to the specific optimization strategy, superior minima may remain accessible. A natural strategy is to relax the locality constraint by extending the search to $k$-local Clifford gates ($k > 2$). It is important to distinguish this approach from merely increasing the depth $d$ of a $2$-local circuit. Given the greedy nature of the algorithm, an optimization restricted to $(k-1)$-local updates may become trapped in local minima that a direct $k$-local search would bypass. Consequently, we treat locality and depth as distinct, complementary resources. In exploring extended $k$-local terms, our primary objective is to elucidate the fundamental limits to disentangling efforts rather than to propose a scalable heuristic, as the search space grows prohibitively large for practical applications.

We begin by analyzing the efficacy of $3$-qubit Clifford disentanglers. Although the full group $\mathcal{C}_3$ contains a staggering $92897280$ elements~\cite{PhysRevLett.122.200502}, this complexity is largely redundant for entanglement cooling. Following the reasoning employed for the $2$-local case, we invoke the local unitary invariance of entanglement to prune the search space. By factoring out the subgroup of local operations, i.e., the tensor product of three single-qubit Clifford unitaries $\mathcal{C}_1^{\otimes 3}$ acting on the individual sites, we collapse the group into distinct equivalence classes. This reduction yields a manageable set of $6720$ canonical representatives, each possessing unique entangling capabilities.

\begin{figure}[h]
	\centering
	\includegraphics[width=0.95\linewidth]{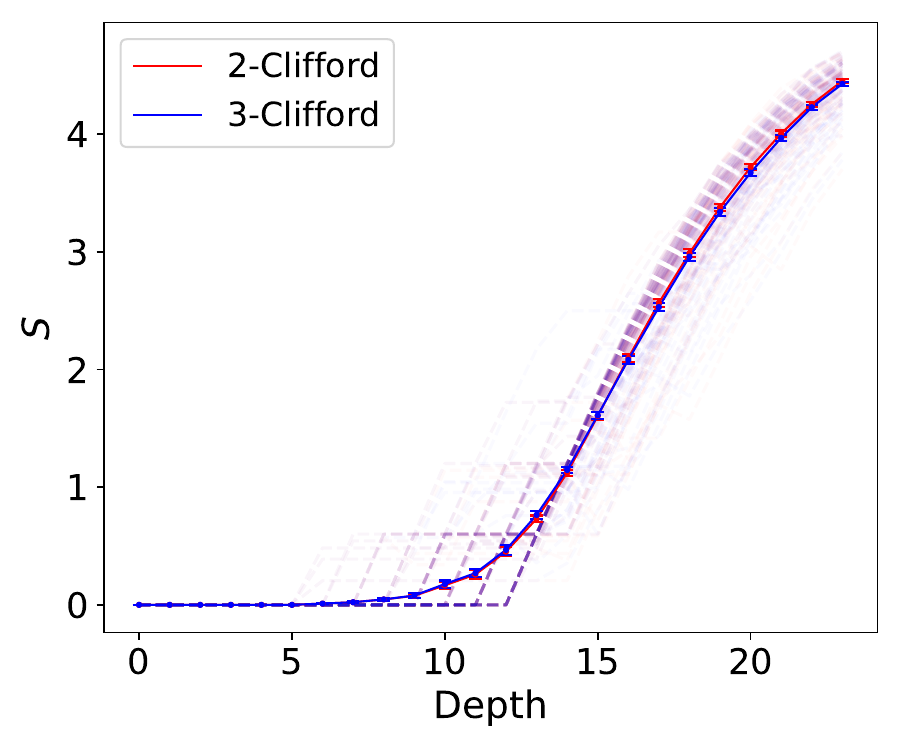}
	\caption{Comparison between the average (solid lines) entanglement of $m=100$ circuits using heuristic optimization with $k=2$ (red) vs $k=3$ (blue) on a sTN simulating a circuit of $12$ qubits with up to $24$ $T$-gates. In dashed, partially transparent lines we show each individual run, whereas the error bars display the standard error of the mean $\sigma/\sqrt{m}$.}
	\label{fig:ent_cool_2v3}
\end{figure}

We implement the optimization by performing sweeps across all contiguous three-site clusters. As mentioned earlier, the optimization objective for $k>2$ is not well determined from the $k=2$ case. In our case, we restrict that entanglement may not increase in any bond in the cluster, and it must improve at least in one of the bonds. For each triplet, we exhaustively evaluate the action of $6720$ representatives, computing the entanglement entropy across the internal bonds to identify the optimal disentangler. Unlike a 'first-improvement' heuristic, this exhaustive search ensures the best local move is selected at each step. While the optimization itself is deterministic, the final cooling efficiency depends heavily on the specific realization of the random circuit, including the sequence of Clifford gates and the spacetime locations of the non-Clifford rotations. To mitigate this variance, we report results averaged over an ensemble of circuit realizations. Similarly to the example in Fig.~\ref{fig:ent_cool_example}, we consider states prepared by circuit involving interspersed action of global $N$-qubit Clifford unitary and a $T$-gate. Although the computational cost of the $k=3$ protocol is higher than its $k=2$ counterpart, we perform a comparative benchmark on small-scale systems ($N=12$) to quantify potential advantages, as shown in Fig.~\ref{fig:ent_cool_2v3}. We clearly observe no improvement of the 3-local entanglement cooling over the 2-local procedure.

\begin{figure}[h]
	\centering
	\includegraphics[width=\linewidth]{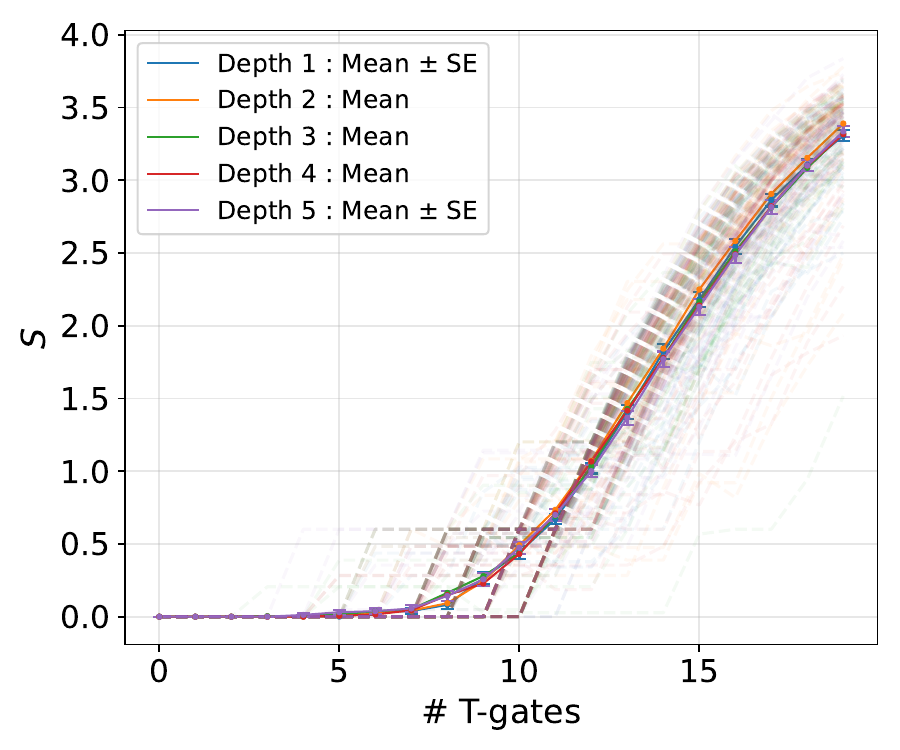}
	\caption{Entropy (y axis) of a state constructed with Clifford and $x$ T-gates (x axis), after being disentangled by the $2$-local heuristic with different sweep depths $d$, in different colors. The error bars represent the standard error (SE) of the mean ($\sigma/\sqrt{N}$) for the two extreme cases (depth 1 and depth 5).}
	\label{fig:depth_exp}
\end{figure}

The effect of increasing the depth can also be analyzed independently. Repeating the disentangling experiment above using $2$-Cliffords, we see in Fig.~\ref{fig:depth_exp} that there is not a clear advantage in increasing depth. Since the optimization is restricted to using a local-sweeping approach, we cannot extrapolate the disentangling achieved as the optimal over the whole $n$-Clifford group. In a similar vein, this structure also does not guarantee that the optimization with additional depth can reach the same result as that with increased locality. Regardless, at least in this regime, the results are consistent and point to both $2$- and $3$-local approaches having the same disentangling power.

Clifford compilation offers a potent tool for dissecting the mechanics of entanglement cooling in these regimes. Notably, any Clifford operation can be synthesized as a sequence of single-type gate layers (e.g. the sequence defined in~\cite{bravyiHadamardfreeCircuitsExpose2021} consisting of $10$ layers in the shape H-CX-P-CX-P-CX-H-P-CX-P). Ongoing refinements to these decomposition algorithms~\cite{Maslov_2023} hold the potential to clarify the relationship between circuit depth and disentangling power, offering a structured alternative to local greedy searches.


\subsection{Optimality of 2-local heuristic}\label{sec:exact_equals_heur}

\begin{figure}[!ht]
	\centering
	\includegraphics[width=\linewidth]{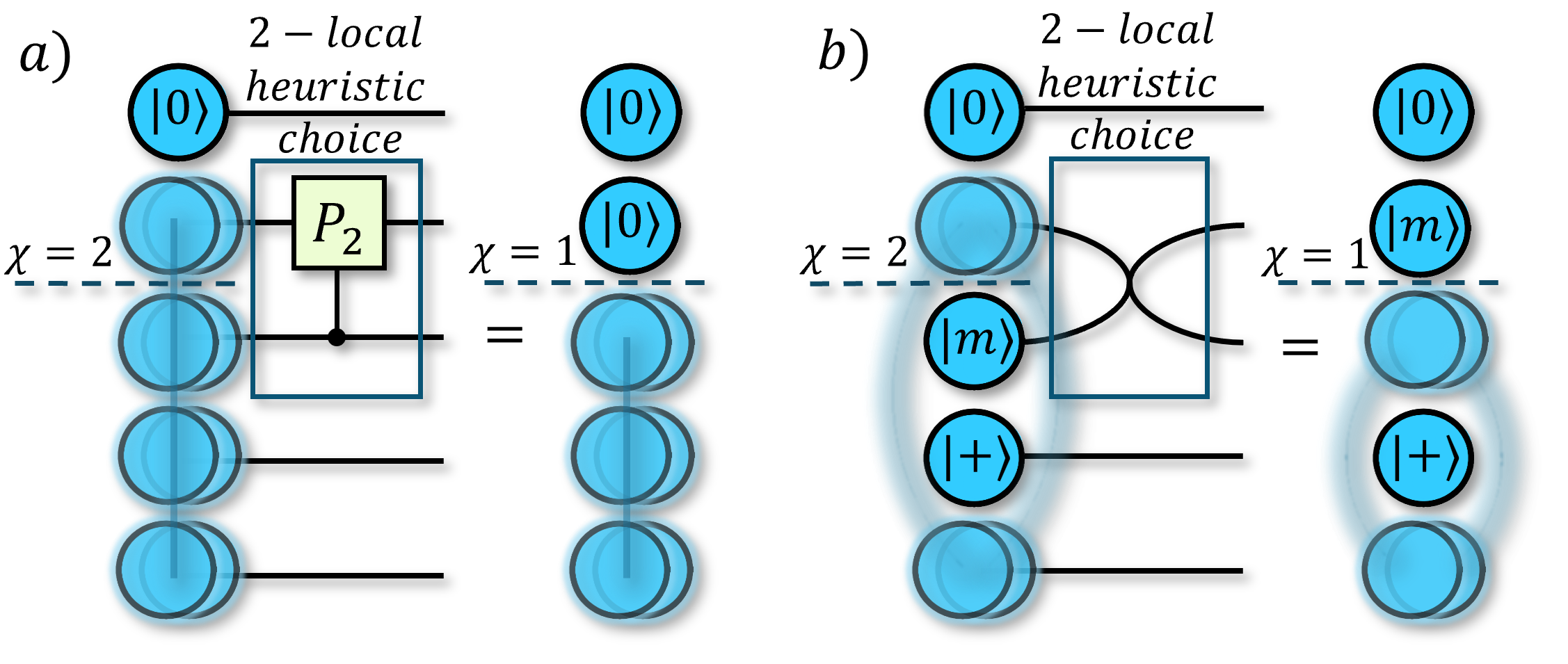}
	\caption{Entanglement structure of the TN in a CTN after applying a conjugated rotation such as that in Fig. \ref{fig:ham_ev_cmps}-a)-ii). A diffuse line represents entangled qubits in a superposition. Both in cases a) and b), we can always find a candidate in $\mathcal{C}_2$ that improves the bond dimension $\chi$ of the MPS and therefore the entropy $S$.}
	\label{fig:exact_ec_exp}
\end{figure}

Elucidating the physical mechanism that enables a local greedy search to reproduce the results of exact entanglement cooling in the $T<N$ regime is of central importance. A priori, the success of such a simple heuristic is counterintuitive given the vastness of the search space; however, it becomes plausible when considering the rigid algebraic structure of the Clifford group. Our analysis reveals that the efficiency of the method arises precisely because the interactions are $2$-local, rather than despite this constraint. To understand this, we consider the state generated by applying a Clifford-conjugated $\mathcal{T}$-gate $U_C T U_C^{\dagger} = \left(  \alpha \: I + \beta\: P_1 \dots P_n \right)$ to a separable product state $\bigotimes_i \ket{\psi_i}$. In a one-dimensional MPS geometry, this operation generates entanglement between distant qubits only in specific configurations, as evidenced by the structure:
\begin{equation}
    \left(  \alpha \: I + \beta\: P_1 \dots P_n \right) \ket{\psi_1}\otimes\dots \otimes\ket{\psi_n}.
\end{equation}
As illustrated in Fig.~\ref{fig:exact_ec_exp}, the two-qubit Clifford group $\mathcal{C}_2$ always contains a gate capable of decoupling the qubit nearest to the boundary (up to local basis changes).  In scenario (a) of the exact entanglement cooling, the optimal disentangler corresponds to a controlled-Pauli rotation (where the control basis is fixed by the stabilizer), while in scenario (b), the optimal operation is a SWAP gate.  By applying this logic iteratively inwards from the edges, one systematically recovers the exact disentangling circuit, a procedure we detail in Appendix~\ref{app:exact_dis} and schematically depict in Fig.~\ref{fig:ent_cool_pre_exact}.

Since the successful application of a disentangler restores the state to a product form, the procedure effectively resets the local environment. This allows the same disentangling logic to be recycled iteratively throughout the chain.


\subsection{Limits on exact entanglement cooling}\label{sec:ent_cool_limits}

The prospect of generalizing exact entanglement cooling to arbitrary input states represents a highly attractive direction for CTN research. Nevertheless, we show that this goal faces hard theoretical barriers. We provide a proof that, when restricted to the Clifford group, there exist fundamental bounds on the types of entanglement that can be successfully removed.

We first investigate the potential for generalizing the exact disentangler beyond stabilizer states. A natural candidate is a heuristic approach applicable to separable input states $\bigotimes_{i=1}^n \ket{\phi_i}$ that do not necessarily lie within the stabilizer manifold. To maintain consistency with previous notation, we consider a non-Clifford rotation $R = \alpha \mathbb{I} + \beta P^0_i$ acting on site $i$. Upon conjugation by the Clifford operation $\mathcal{C}$, this operator maps to a global rotation $R_\mathcal{C} = \alpha \mathbb{I} + \beta P$, where the coefficients $\alpha, \beta$ are preserved but the local Pauli $P^0_i$ transforms into a multi-qubit Pauli string $P = P_1 \otimes \dots \otimes P_n$. To ensure the operation is non-trivial, we assume $P$ has a weight of at least two (i.e., $P_n \neq \mathbb{I}$ and $P_j \neq \mathbb{I}$ for some $j < n$), as otherwise the gate is non-entangling. If we apply $R_\mathcal{C}$ to an arbitrary product state $\bigotimes_{i=1}^n \ket{\phi_i}$, we cannot expect to consistently disentangle the result via Clifford operations. If such a disentangling were universally possible, the output would be a new product state $\bigotimes_{i=1}^n \ket{\tilde{\phi}_i}$; this would allow for the iterative simulation of arbitrary quantum circuits with polynomial resources, implying an efficient classical simulation of universal quantum computation.

Given the impossibility of universal global disentangling, we turn to the immediately weaker requirement: disentangling a single qubit. We consider a generic initial state $\ket{\Psi} \otimes \ket{\phi}_n$, where the $n$-th qubit is separable but the remainder is arbitrary. We seek a unitary $\mathcal{U}$ such that:
\begin{equation}\label{eq:cool_limits}
	\mathcal{U}\left(  \alpha I + \beta P_1 \dots P_n \right) \ket{\Psi}\otimes\ket{\phi_n} = \ket{\tilde{\Psi}}\otimes\ket{\tilde{\phi}_n},
\end{equation} 
and we investigate whether this condition can be satisfied by a Clifford unitary. Without loss of generality, we designate the last qubit as the target; since the Clifford group contains the SWAP gate, a solution for any site $k$ implies a solution for site $n$ (and vice versa) without altering the complexity class of $\mathcal{U}$. 
Our statement is then the following:
\begin{theorem}\label{thm:no_ec_improvement}
	Let be an $N$-qubit rotation
    \begin{equation}\label{eq:theorem_unitary}
        e^{-i\theta P_1 \dots P_n} = \alpha I + \beta P_1 \dots P_n
    \end{equation}
    with $\alpha=\cos(\theta),\beta=-i\sin(\theta)$, and an initial state 
    \begin{equation}\label{eq:theorem_state}
        \ket{\Psi}\otimes\ket{\phi_n}
    \end{equation}
    where $\ket{\Psi}$ is an $N-1$ qubit state and $\ket{\phi_n}$ is a $1$-qubit state. If a unitary $\mathcal{U}$ is applied to $\left(  \alpha I + \beta P_1 \dots P_n \right) \ket{\Psi}\otimes\ket{\phi_n}$, for arbitrary $\theta$,$\ket{\Psi}$, that results in a state of the form $\ket{\tilde{\Psi}}\otimes\ket{\tilde{\phi}_n}$ (\ie the last qubit is separable), then $\mathcal{U}$ is in the Clifford group if and only if $\ket{\phi_n}$ is a stabilizer state.
\end{theorem}

In other words, there are no \textit{complete} disentanglers beyond the known exact one. At a high level, the proof proceeds by analyzing the output state $\mathcal{U} \; e^{-i\theta P_1 \dots P_n} \; \ket{\Psi}\otimes\ket{\phi_n}$ using a decomposition of $\calU$ over the states of a single qubit $\mathcal{U} = \mathcal{U}_1\otimes \ket{\omega}\bra{\phi_n} + \mathcal{U}_2\otimes \ket{\bar{\omega}}\bra{\bar{\phi}_n}$, using some conveniently chosen states that depend on $\mathcal{U}$ and the other inputs. After tracing out the first $n-1$ qubits from such output, we can translate the hypothesis of \textit{the output being separable} into a purity condition on the reduced density matrix of the $n$-th qubit. This constraint imposes a rigid structure on $\mathcal{U}$. 
We then test this structure against a specific input-dependent stabilizer state.
We demonstrate that if the input $\ket{\phi_n}$ is not a stabilizer state, the output inevitably fails to remain a stabilizer state, thereby violating the Clifford preservation property. Conversely, if $\ket{\phi_n}$ is a stabilizer state, the known exact disentangler applies. This establishes the "if and only if" statement of the theorem. The details are provided in Appendix \ref{app:proof}.


\section{Entanglement build-up with non-Clifford rotations}\label{sec:ent_exp}

While entanglement cooling improves CTN performance across the board, it is instructive to analyze settings that play to the specific advantages of tensor networks. Standard quantum compilation favors the Clifford+$T$ set, a choice dictated by the requirements of quantum error correction and magic state preparation. Crucially, however, the tensor network part of the ansatz, $\mathcal{T}$, is not subject to these hardware limitations. Because $\mathcal{T}$ encodes the state vector directly, it handles arbitrary non-Clifford rotations with equal ease, allowing us to bypass the artificial discrete structure imposed by the Clifford+$T$ paradigm.

The introduction of arbitrary rotation angles alters the interpretation of the bond dimension $\chi$ compared to standard Tensor Networks. In the CTN framework, $\chi$ does not merely bound the bipartite entanglement; it also serves as a proxy for non-stabilizerness, in the sense that $\chi > 1$ indicates a superposition of multiple stabilizer states. Consequently, the resource cost of a simulation becomes a continuous function of the rotation angle $\theta$. Analogous to how a weakly entangling gate requires minimal bond dimension, a non-Clifford rotation close to the identity introduces only a small amount of magic. This continuity suggests that compiling circuits into Clifford gates plus arbitrary continuous rotations is advantageous, yielding significantly shallower circuits than standard Clifford+$T$ decomposition. It is therefore useful to quantify the precise scaling behavior of entanglement and $\chi$ as a function of the rotation parameters.

\begin{figure}[!ht]
	\centering
	\includegraphics[width=\linewidth]{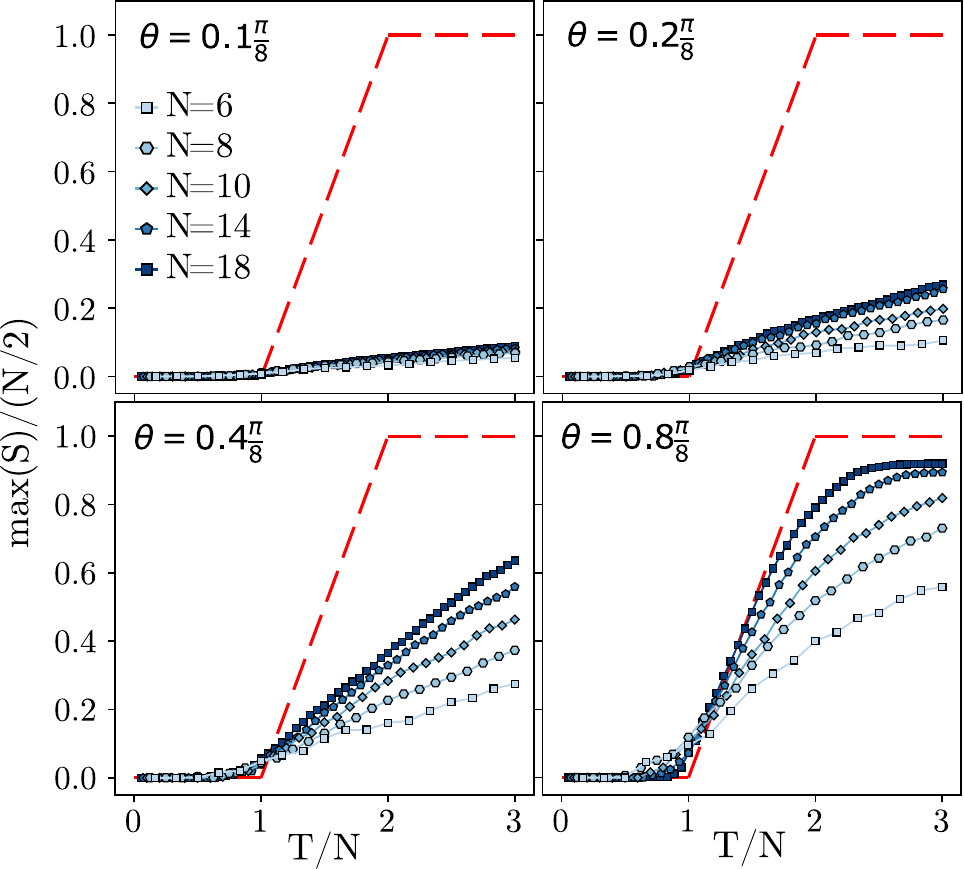}
	\caption{Entropy of the equipartition of an CTN as the number of rotations in the circuit for the given angle increase.}
	\label{fig:ent_rotations}
\end{figure}

Repeating the simulation from Fig.~\ref{fig:ent_cool_example} with smaller rotation angles allows us to see this behavior. Using a $T$-gate rotation introduces as much magic as possible for a single qubit rotation. Notice that the $T$ gate is equivalent to a $Z$ rotation with angle $\pi/4$, that is:
\begin{eqnarray}\label{eq:tgate_long}
	T &=& \begin{pmatrix} 1 & 0 \\ 0 & e^{-i\frac{\pi}{4}} \end{pmatrix} \simeq \begin{pmatrix} e^{i\frac{\pi}{8}} & 0 \\ 0 & e^{-i\frac{\pi}{8}} \end{pmatrix} \\ &=&  \cos\left(\frac{\pi}{8}\right) I -i \sin\left(\frac{\pi}{8}\right) Z = e^{-i\frac{\pi}{8}Z} = R_Z(\pi/4). \nonumber
\end{eqnarray}
Rotations with smaller angles are expected to generate entanglement more gradually. Figure~\ref{fig:ent_rotations} displays the entropy growth as we increase the number $t$ of non-Clifford rotations for different system sizes. The simulation utilizes random Clifford layers~\cite{bravyiHadamardfreeCircuitsExpose2021} and random local $R_Z(\theta)$ gates. At each step, we minimize the bond dimension using the exact cooling protocol (Prop.~\ref{prop:exact_EC}) whenever the state structure permits, and the $2$-local heuristic (Def.~\ref{def:heur_EC}) otherwise.

The three characteristic regimes of entanglement cooling identified previously reappear in Fig.~\ref{fig:ent_rotations}, albeit with thresholds shifted along the $x$-axis (gate count). Notably, for sufficiently small angles, the second threshold is pushed beyond the observation window. The initial phase ($T < N$) is consistent with our earlier analysis: the exact heuristic remains highly effective, suppressing entanglement growth and defining a 'classical' regime where CTN simulation is efficient. Beyond this critical point, linear entropy scaling re-emerges. This accumulation arises from the increasing failure rate of the exact disentangling scheme (Fig.~\ref{fig:ent_cool_exact}); averaging over the ensemble reveals a growing fraction of trajectories where complete disentanglement is impossible. Crucially, however, the growth rate is proportional to the rotation angle
, which consequently delays the saturation threshold. This has significant implications for variational quantum circuits: the domain of efficient simulation extends from $N$ $T$-gates to a depth inversely proportional to the average rotation angle. Specifically, if the parameters cluster around $\bar{\theta} \sim \frac{1}{\gamma} (\pi/4)$, it requires approximately $\gamma$ rotations to increment the bond entropy by one unit.

\begin{figure}[h]
	\centering
	\includegraphics[width=0.7\linewidth]{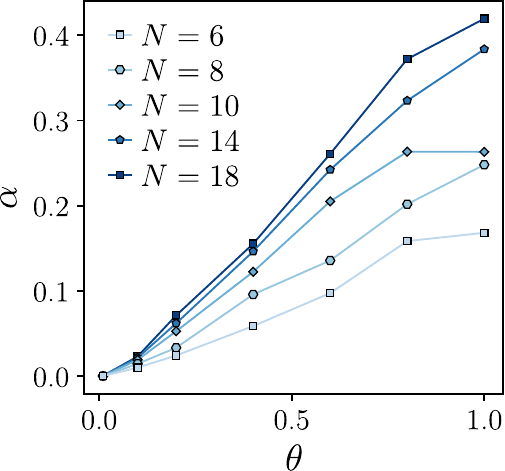}
	\caption{Growth of entropy per $T$-gate for different sizes of the circuit.}
	\label{fig:ent_slopes}
\end{figure}

To accurately quantify the dynamics of the intermediate regime ($N < T < 2N$), we introduce the entropy growth rate $\alpha = \Delta S / N$, defined as the total entropy accumulation $\Delta S$ normalized by the number of non-Clifford gates in this interval. This metric captures the average entanglement cost per gate in the transition zone between exact disentangling and saturation.  As shown in Fig.~\ref{fig:ent_slopes}, $\alpha$ scales linearly with the rotation angle $\theta$. This proportionality is advantageous for practical simulation: it implies that for small-angle rotations, the effective 'faithful depth' of the CTN ansatz extends well beyond the nominal count of non-Clifford gates, as the cost per gate is reduced. However, from a theoretical standpoint, the result aligns strictly with expectations, revealing no anomalous low-entanglement subspaces or exploitable non-linearities beyond this standard resource scaling.

\begin{figure}[ht!]
	\centering
	\includegraphics[width=\linewidth]{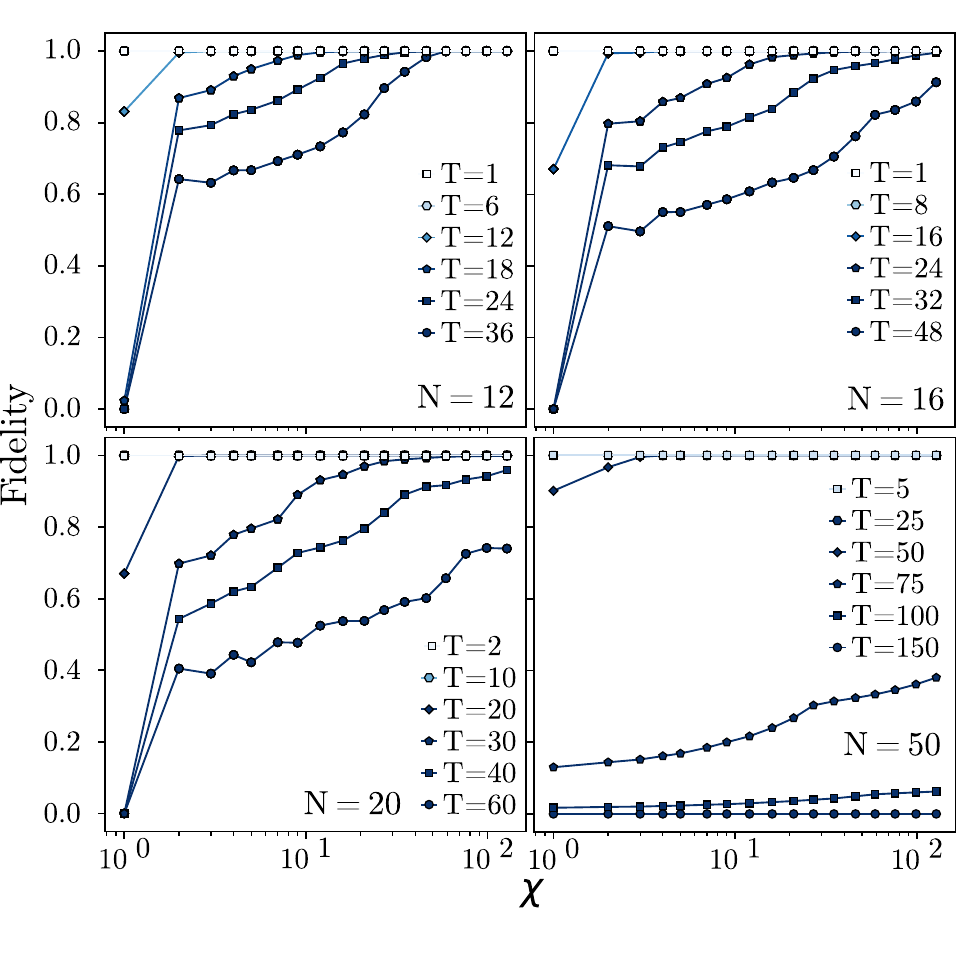}
	\caption{Fidelity over the inverse of bond dimension, for $N=12,16,20,50$.}
	\label{fig:ent_fidelity}
\end{figure}

The approximation of a quantum state $\ket{\psi}$ with a TN $\ket{\psi_\chi}$ using bond dimension $\chi$ is useful whenever the fidelity $F=|\braket{\psi | \psi_\chi}|^2$ between them is $F\sim 1$. The scaling of $F$ with bond dimension $\chi$ is thus a critical metric for benchmarking the CTN ansatz as an extension of TN methods. Particularly, because $\chi$ parametrizes the non-stabilizerness of the state rather than simple bipartite entanglement, and consequently standard convergence criteria from MPS do not automatically apply. To characterize this relationship, we simulate the resource buildup via random $T$-gate injection and analyze the fidelity as a function of $\chi$ (Fig.~\ref{fig:ent_fidelity}). As expected for generic quantum states, high fidelities are inaccessible at low bond dimensions; however, the convergence is monotonic. Crucially, the profile lacks the steep initial jumps associated with states governed by a rapidly decaying singular value spectrum. Instead, we observe a gradual accumulation of fidelity, indicative of a 'flat' entanglement spectrum where information is delocalized across many modes, a hallmark of highly non-stabilizer states. For larger systems ($N \geq 16$), computational limits prevent reaching unity fidelity, yet the scaling behavior remains consistent with smaller instances.

\section{Conclusions}
\label{sec:conc}
In this work, we investigate the capabilities and limitations of Clifford-based disentangling strategies within tensor network simulation frameworks. Our main conclusions can be summarized as follows. First, we show that existing heuristic disentanglers cannot easily be improved with an increase on depth, or with a larger dimension of the Clifford space used in the local update step. In parallel, we prove for exact heuristics that there does not exist a more general algorithm that completely disentangles even a single qubit. Second, effective disentangling implies that quantum circuits composed of 'Clifford + T gates' with linear depth in the number of qubits can be simulated with very low classical cost within the Clifford-augmented tensor network framework. In this regime, entanglement cooling prevents the rapid growth of tensor-network resources and enables efficient classical tractability. Building on our analysis of resource growth as a function of rotation angle, we find that the simulability of variational quantum circuits can be substantially extended when considering the average magnitude of rotation angles rather than their mere count. Small-angle non-Clifford rotations introduce magic gradually, on average, allowing the efficient simulation regime to persist to depths well beyond those expected from discrete gate-count arguments. More generally, disentangling proves to be a powerful tool for extending tensor network–based simulation methods beyond their conventional limits. At the same time, our results provide a quantitative bound on the extent of its effectiveness, clarifying the regimes in which Clifford-based disentangling can and cannot mitigate the accumulation of non-Clifford resources.

Looking forward, an important direction for future work is the development of improved disentangling algorithms. In particular, it would be valuable to identify systematic procedures for partial disentangling that either expose the internal structure of heuristic outputs or enhance their performance in regimes where full disentangling is impossible.
While in this work we focused on MPS in one-dimensional system, a natural directions for further work is to considered enhancing with Clifford operations other classes of tensor network states including tree tensor networks~\cite{Tagliacozzo09, Evenbly09} or tensor networks in higher dimensional systems~\cite{Jordan08, Orus14}, including new disentangling heuristics that are tailored to such structures.
Further extensions of this work concern the generalization of entanglement cooling techniques to resources beyond non stabilizernes, such as non-Gaussianity in fermionic~\cite{Sierant26ferm, Wu25disent, Bittel25optimal} and bosonic~\cite{Chabaud20stellar, Girardi25testing} systems. Exploring how similar ideas apply to other notions of quantum complexity may further broaden the scope of classical simulation strategies for quantum many-body systems.


\section*{Acknowledgements}
\label{sec:ack}
We thank Neil Dowling, Xhek Turkeshi, Guglielmo Lami, Emanuele Tirrito and Mario Collura for helpful discussions. 
The authors acknowledge support from EU Grant No. HORIZON-EIC-2022-PATHFINDEROPEN-01-101099697, QUADRATURE.  This
work is also financially supported by the Ministry
of Economic Affairs and Digital Transformation of
the Spanish Government through the QUANTUM
ENIA project call – Quantum Spain project, and
by the European Union through the Recovery,
Transformation and Resilience Plan – NextGenerationEU within the framework of the Digital Spain
2026 Agenda.
P.~Sierant acknowledges fellowship within the “Generación D” initiative, Red.es, Ministerio para la Transformación Digital y de la Función Pública, for talent attraction (C005/24-ED CV1), funded by the European Union NextGenerationEU funds through PRTR.

\bibliography{references}
\nocite{*}

\onecolumngrid


\appendix


\section{Exact entanglement cooling}\label{app:exact_dis}

\begin{figure}[h]
	\centering
	\includegraphics[width=0.72\linewidth]{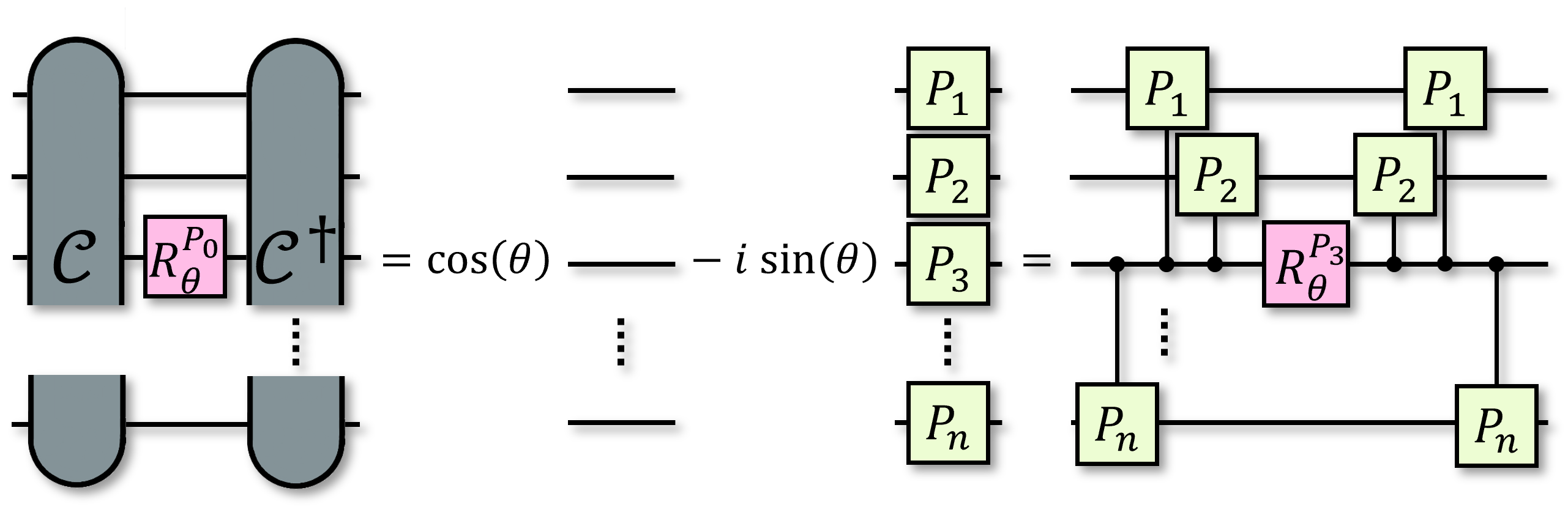}
	\caption{Outcome of commuting a Pauli operator $P_0$ with an arbitrary Clifford circuit $\mathcal{C}$, and a possible decomposition into a circuit. This example assumes $P_i$ on site $3$ is not trivial, otherwise the same decomposition holds centered on a different qubit. If $P_3=Z$, then the circuit also needs, on site $3$, $H$ gates around the rotation, at the beginning, and at the end.}
	\label{fig:ent_cool_pre_exact}
\end{figure}

In this section, we present in our language how the exact entanglement cooling algorithm \cite{Fux25disent,exact_lami,Liu25cliff} works. Let us first point out that for a given CTN, the equations for circuit simulation (described in Ref.~\cite{stabTN}) that relate to the situation in Fig. \ref{fig:ham_ev_cmps}-a)-ii) where an arbitrary rotation in the $Z$ axis is processed through $\calC$, amount to the equivalency in Fig.~\ref{fig:ent_cool_pre_exact}.

In the figure, we assume that $P_3$ is either $X$ or $Y$. Otherwise, if it is $Z$ then we must add $H$ gates before and after each sequence of ``controls", and if it is the identity we must choose as the centre of the controlled operations a different qubit (there must be one unless it is a trivial rotation). In this situation, $P=P_1\dots P_n$ is an arbitrary Pauli string. Since the compilation into Clifford + local rotations produces this kind of global rotation, the exact entanglement cooling stated in the main text and restated below becomes very useful:

\begin{definition*}[Exact entanglement cooling]
For a CTN state $(\mathcal{C}, \mathcal{T})$ representing $n$ qubits, in which the tensor network component $\mathcal{T}$ factorizes such that qubit $i$ is in a single-qubit stabilizer state $|\sigma\rangle_i$ (an eigenstate of a Pauli operator $P_c$), let $U = \alpha \mathbb{I} + \beta P$ be a rotation generated by the Pauli string $P = \bigotimes_{k} P_k$. If the local Pauli operator $P_i$ acts non-trivially on the state (i.e., $P_i |\sigma\rangle_i \neq |\sigma\rangle_i$), then the global operation $U$ can be perfectly disentangled. Specifically, $U$ decomposes into a local rotation $\alpha \mathbb{I} + \beta P_i$ acting solely on qubit $i$, and a cascade of controlled-Pauli gates absorbed into the Clifford operation $\mathcal{C}$. These entangling gates are controlled by qubit $i$ in the $P_c$ basis, applying the corresponding Pauli $P_j$ to each target qubit $j \neq i$.
\end{definition*}

\begin{figure}
	\centering
	\includegraphics[width=0.64\linewidth]{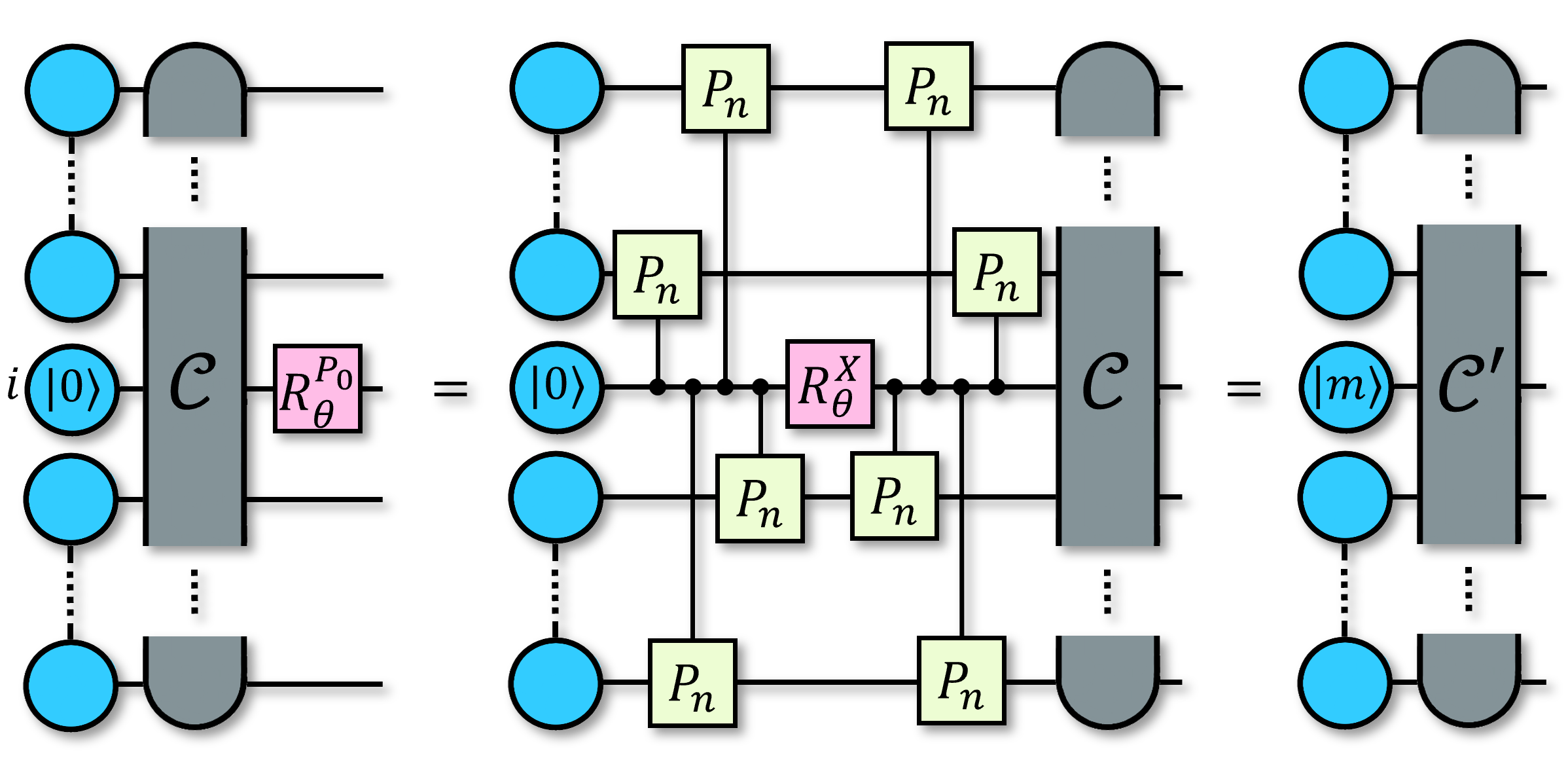}
	\caption{Example of the exact disentangling procedure. An arbitrary rotation, which has magic, is mapped to a global rotation on the TN part of the CTN. With a clever decomposition, it can be turned back again into a local rotation, thus not increasing $\chi$ of the TN.}
	\label{fig:ent_cool_exact}
\end{figure}

We illustrate why it works at a high level in Fig.~\ref{fig:ent_cool_exact} with an example using $\ket{\sigma}=\ket{0}$ on qubit $i$ and a Pauli string with $P_i=X$ (for $\ket{0}$ we could have $P_i\in\{X,Y\}$). The controlled gates to the left of the central rotation (in pink) act trivially on qubit $i$, whereas the ones on the right, as they are a Clifford gate, can be absorbed into $\mathcal{C}$ resulting in a different $\mathcal{C}'$. Hence, we are left with a local rotation acting on the MPS part (or TN, in general), that turns the stabilizer into a magic state. In our example, this is
\begin{equation}\label{eq:m_state}
    \ket{m}= \cos(\pi/8) \ket{0} - i \sin(\pi/8) \ket{1}.
\end{equation}
Any other situation that still fulfils the conditions, i.e. $P$ affects non-trivially any site that is in a separable stabilizer state, can be mapped to our example with simple changes of basis, resulting in a different state in site $i$ that will regardless always contain some magic. This is consistent with the recipe at the end of Prop.~\ref{prop:exact_EC} for the list of controlled gates.

This algorithm works because the starting state is by assumption a separable state in the computational basis, so it fits the requirement, and every time we apply it on a state with $m$ separable stabilizer states in $\mathcal{T}$, it returns a new $\mathcal{T}'$ with $m-1$ of such states. Random Cliffords, on average, affect many qubits. This makes it more likely than not that we can implement this disentangler after arbitrary rotations, but as $m$ decreases it gets statistically less likely that Prop.~\ref{prop:exact_EC} works, with an upper limit of $n$ number of applications after which all the sites have been ``used up". This setup can also be described in terms of error correction codes, as shown in \cite{Fux25disent}. It also explains why we can disentangle so much up to $n$ $T$-gates. After we have applied this many, there are no known approaches that work as well as the local disentangler. In fact, we prove in the next section that we cannot have a universal formula that disentangles at least one qubit completely, like the one we explained does.


\section{Complete proof for the No cooler disentanglement}\label{app:proof}

 In this section we give the complete proof of Theorem \ref{thm:no_ec_improvement}. To do so, we use that for an arbitrary state $\ket{\phi_n}$, we can write any unitary as
\begin{equation}\label{eq:u_decomp}
	\forall \mathcal{U},\ket{\phi_n} \; \exists \: \mathcal{U}_1, \mathcal{U}_2 \; \text{s.t.} \; \mathcal{U} = \mathcal{U}_1\otimes \ket{\omega}\bra{\phi_n} + \mathcal{U}_2\otimes \ket{\bar{\omega}}\bra{\bar{\phi}_n},
\end{equation}
where $\ket{\omega}$ is an arbitrary state that depends on $\mathcal{U}$ and $\ket{\bar{\phi}_n}$ ($\ket{\bar{\omega}}$) is orthogonal to $\ket{\phi_n}$ ($\ket{\omega}$). The individual steps are easier to understand if we first allow some extra assumptions that simplify the equations, which we remove later to prove the general case.

\subsection{Simplifed proof}
Let us make the assumptions $\alpha=\cos(\pi/8),\beta= -i \sin(\pi/8)$ and $P_n \ket{\phi_n} = \ket{\bar{\phi}_n}$. This is the case if the non-Clifford $R$ is a $T$-gate, for the former, and when $\ket{\phi_n}=\ket{m}$ (Eq. \ref{eq:m_state}) with $P_n=X$ for the latter. Specifically:
\begin{equation}\label{eq:case_1}
	\ket{\phi_n} = \ket{m} = \cos(\pi/8)\ket{0} -i \sin(\pi/8)\ket{1} = \alpha\ket{0} +\beta\ket{1} \; , \; \ket{\bar{\phi}_n} = \ket{\bar{m}} = \beta\ket{0} + \alpha\ket{1}
\end{equation}
Then, the final state is:
\begin{equation}\label{eq:u_decomp_c1}
	\left( \mathcal{U}_1\otimes \ket{\omega}\bra{m} + \mathcal{U}_2\otimes \ket{\bar{\omega}}\bra{\bar{m}} \right) \left(  \alpha I + \beta P \right) \ket{\Psi}\otimes\ket{m} = \ket{\tilde{\Psi}}\otimes\ket{\tilde{\phi}_n}.
\end{equation}
Tracing out the first $n-1$ qubits (which we call subsystem $B$) on the corresponding density matrix $\rho_f$ should give us a pure state, since by hypothesis the final state is separable. Using the notation $P_B=P_1\dots P_{n-1}$, we get
\begin{equation}\label{eq:trace_init_c1}
\begin{split}
	&\text{Tr}_B\left[ \rho_f \right] = \text{Tr}_B \left[ \left( \mathcal{U}_1\otimes \ket{\omega}\bra{m} + \mathcal{U}_2\otimes \ket{\bar{\omega}}\bra{\bar{m}} \right) \left( \alpha \ket{\Psi}\otimes\ket{m} + \beta P_B\ket{\Psi}\otimes\ket{\bar{m}} \right) \cdot \right. \\ 
	& \qquad \left. \cdot \left( \alpha^*  \bra{\Psi}\otimes\bra{m} + \beta^* P_B\bra{\Psi}\otimes\bra{\bar{m}} \right) \left( \mathcal{U}^\dagger_1 \otimes \ket{m}\bra{\omega} + \mathcal{U}^\dagger_2\otimes \ket{\bar{m}}\bra{\bar{\omega}} \right) \right] = \\
	= \text{Tr}_B &\left[ \left( \alpha \left( \mathcal{U}_1 \ket{\Psi} \right) \otimes \bra{\omega} + \beta \left(\mathcal{U}_2 P_B\ket{\Psi} \right)\otimes \ket{\bar{\omega}} \right) \left(\alpha^* \left(\bra{\omega} \mathcal{U}^\dagger_1 \right) \otimes \bra{\omega} + \beta^* \left(\bra{\Psi}P_B\mathcal{U}^\dagger_2 \right) \otimes \bra{\bar{\omega}} \right) \right] \equiv \\
	& \quad \equiv \text{Tr}_B \left[ \left(\alpha \ket{\Omega_1} \otimes \bra{\omega} + \beta \ket{\Omega_2}\otimes \ket{\bar{\omega}} \right) \left(\alpha^* \bra{\Omega_1} \otimes \bra{\omega} + \beta^* \bra{\Omega_2} \otimes \bra{\bar{\omega}} \right) \right] .      
\end{split} 
\end{equation}
In the last step, we renamed $\mathcal{U}_1 \ket{\Psi} \equiv \ket{\Omega_1}$ and $\mathcal{U}_2 P_B \ket{\Psi} \equiv \ket{\Omega_2}$, which we can use to make an orthonormal set of states:
\begin{equation}\label{eq:basis_c1}
	\ket{\tilde{\Omega}_1}= \ket{\Omega_1} \quad , \quad \ket{\tilde{\Omega}_2}=\frac{\ket{\Omega_2} - \braket{\Omega_1 | \Omega_2} \ket{\Omega_1}}{\sqrt{1+\braket{\Omega_1 | \Omega_2}^2}}
\end{equation}
The trace can be calculated with any arbitrary completion of this set in the subspace $B$, such that
\begin{equation}\label{eq:trace_c1}
\begin{split}
	\text{Tr}_B\left[ \rho_f \right] = \bra{\tilde{\Omega}_1} \rho_f\ket{\tilde{\Omega}_1} + \bra{\tilde{\Omega}_2}&\rho_f\ket{\tilde{\Omega}_2} = |\alpha|^2 \ket{\omega}\bra{\omega} + \alpha^* \beta  \braket{\Omega_1|\Omega_2}\ket{\bar{\omega}}\bra{\omega} + \\ + \alpha \beta^*  \braket{\Omega_2|\Omega_1}\ket{\omega}\bra{\bar{\omega}}& + |\beta|^2 \ket{\bar{\omega}}\bra{\bar{\omega}} + \left( \frac{1-\braket{\Omega_2|\Omega_1}^2}{\sqrt{1+\braket{\Omega_2|\Omega_1}^2}} \right)^2 |\beta|^2 \ket{\bar{\omega}}\bra{\bar{\omega}} = \\ = \left(\alpha \ket{\omega} + \beta \braket{\Omega_1|\Omega_2}  \ket{\bar{\omega}} \right) & \left(\alpha^* \bra{\omega} + \beta^* \braket{\Omega_2|\Omega_1}  \bra{\bar{\omega}} \right)  + \left(\frac{1-\braket{\Omega_2|\Omega_1}^2}{\sqrt{1+\braket{\Omega_2|\Omega_1}^2}} \right)^2 |\beta|^2 \ket{\bar{\omega}}\bra{\bar{\omega}}.
\end{split}
\end{equation}
For this to be a pure state, the structure of the terms with $\ket{\omega}\bra{\omega},\ket{\omega}\bra{\bar{\omega}}$ and $\ket{\bar{\omega}}\bra{\omega}$ force the coefficient of $\ket{\bar{\omega}}\bra{\bar{\omega}}$, so the only solution is for the rightmost term in the last line of Eq.\ref{eq:trace_c1} to be $0$. Thus
\begin{equation}\label{eq:pre_condition_c1}
	\left(\frac{1-\braket{\Omega_2|\Omega_1}^2}{\sqrt{1+\braket{\Omega_2|\Omega_1}^2}} \right)^2 |\beta|^2 = 0 \leftrightarrow \braket{\Omega_1|\Omega_2} = \pm 1 \leftrightarrow \bra{\Psi}U_1^\dagger U_2 P_B \ket{\Psi} = \pm 1.
\end{equation}
As we ask this to hold for arbitrary $\ket{\Psi}$, this is equivalent to requiring
\begin{equation}\label{eq:condition_c1}
	\mathcal{U}_1^\dagger \mathcal{U}_2 = I \leftrightarrow \mathcal{U}_2 = \mathcal{U}_1 P_B.
\end{equation}
This leaves the full unitary as
\begin{equation}\label{eq:pre_f_unitary_c1}
	\mathcal{U}_f = \mathcal{U}_1 \otimes I \left( \alpha I \otimes \ket{\omega}\bra{m} + \beta P_B \otimes \ket{\bar{\omega}}\bra{\bar{m}} \right).
\end{equation}
We can finally prove that this cannot be a Clifford gate. Consider a stabilizer state $\ket{s} = \bigotimes_{i=1}^{n-1}\ket{s_i}$ such that $P_i\ket{s_i}=\ket{\bar{s}_i}$ (or $\ket{s_i}$ if $P_i=I$) and $\ket{s_n}=\ket{0}$. As we assumed more than one $P_i$ to be non-trivial, the state $\ket{\bar{s}}=\bigotimes_{i=1}^{n-1}\ket{\bar{s}_i}$ fulfills $\braket{s|\bar{s}}=0$. Applying Eq.~\ref{eq:pre_f_unitary_c1} to this state gives us
\begin{equation}\label{eq:last_step_c1}
	\mathcal{U}_f \bigotimes_{i=1}^{n}\ket{s_i} = \alpha \mathcal{U}_1\ket{s} \ket{\omega} + \beta \mathcal{U}_1\ket{\bar{s}} \ket{\bar{\omega}} = (\alpha \: \mathcal{U}_1 \ket{s} \otimes \ket{\omega} + \beta \: \mathcal{U}_1 \ket{\bar{s}} \otimes \ket{\bar{\omega}}).
\end{equation}
This is not a stabilizer state no matter which state $\ket{\omega}$ or unitary $\mathcal{U}_1$ we use, because it is a superposition of two orthogonal states with magic coefficients. It can also be shown with the reduced matrix of qubit $n$; if this was a stabilizer state, the reduced matrix should be either a pure state or the maximally mixed state (this can be proven using the closed form of stabilizer states \cite{Dehaene_2003,de_Silva_2025}). The coefficients $\alpha,\beta$ ensure that it is neither. We will actually use this in the complete proof. In conclusion, we have that $\mathcal{U}_f$ takes a stablizer state to a non-stabilizer state, so it cannot be a Clifford transformation. $\qed$

Notice that we could have had the gadget depend on $\ket{\Psi}$ after Eq.~\ref{eq:pre_condition_c1}. This goes beyond our proof, as we do not rule out that for very specific states there is something that we can do. For example, if the rest of the state does contain a separable stabilizer state, we could reverse the roles of that site and site $n$ and apply the original algorithm. The generalization that we are looking for is something that, as~\cite{exact_lami,Liu25cliff} does, works agnostically of the rest of the state as long as we can identify one affected qubit with the appropriate behaviour (like being a stabilizer state, in the known heuristic). Needing information of the state to decide how to disentangle, on the other hand, will be in general expensive, which also justifies not focusing on this setting.

\subsection{General proof}

Here we are just going to repeat the steps of the previous proof with slightly more complicated equations, as we do not have the simplifications above. Instead, $\alpha, \beta$ are arbitrary and do not match the coefficients of the input states, which we write as generic
\begin{equation}\label{eq:state_c2}
	\ket{\phi_n}=\mu\ket{0} + \nu \ket{1} \: ,\: \ket{\bar{\phi}_n}=\nu^*\ket{0} - \mu^* \ket{1}.
\end{equation}
We also do not assume anything over how $P_n$ acts on the state in qubit $n$, meaning that we have $P\ket{\phi_n} = \gamma \ket{\phi_n} + \delta \ket{\bar{\phi}_n}$. To slightly simplify some equations, we do use that we can always write these states with $\mu,\gamma\in \Re$ as we do not care about the global phase. As per the assumptions of the theorem, we also have
\begin{equation}\label{eq:ab_c2}
	\exists \theta \: \text{s.t.} \: \alpha = \cos(\theta) \: , \: \beta= -i\sin(\theta).
\end{equation}
The action of the rotation $R_\mathcal{C}$ on the initial state is then
\begin{equation}\label{eq:u_decomp_c2}
	\left(  \alpha I + \beta \gamma P_B \right) \ket{\Psi}\otimes\ket{\phi_n} + \beta \delta \: \: P\ket{\Psi}\otimes\ket{\tilde{\phi}_n},
\end{equation}
where $P_B=P_1\dots P_{n-1}$. In this instance, after
applying $\mathcal{U}$ our $\rho_f$ is
\begin{equation}\label{eq:trace_init_c2}
	\begin{split}
		\rho_f = \left( \left( \mathcal{U}_1 \left(  \alpha I + \beta \gamma P_B \right)\ket{\Psi} \right) \otimes \ket{\omega} + \beta \delta \left( \mathcal{U}_2 P_B \ket{\Psi} \right) \otimes \ket{\bar{\omega}} \right) \left( \text{h.c.} \right)    
	\end{split} 
\end{equation}
where we omitted the hermitian conjugate as it follows the structure in Eq.~\ref{eq:trace_c1}. To do the trace, we identify the states as
\begin{equation}\label{eq:basis_c2}
	\begin{split}
	\ket{\Omega_1} &= \mathcal{U}_1 \frac{\alpha I + \beta \gamma P_B}{\sqrt{1-|\beta|^2|\delta|^2}} \ket{\Psi} \\ 
	\ket{\Omega_2} &= \mathcal{U}_2 P \ket{\Psi} 
	\end{split}.
\end{equation}
The denominator in $\ket{\Omega_1}$ comes from the following norm:
\begin{equation}\label{eq:norm_c2}
	\begin{split}
	(\alpha I + \beta\gamma P)&(\alpha^*I + \beta^* \gamma^* P^\dagger) = |\alpha|^2+ + |\beta|^2|\gamma|^2 + 2 \text{Re}(\alpha\beta^*\gamma^*) = \\ =& 1 - |\beta|^2 + |\beta|^2|\gamma|^2 = 1 - |\beta|^2(1-|\gamma|^2) = 1 - |\beta|^2|\delta|^2
	\end{split}
\end{equation}
where we used Eq.~\ref{eq:ab_c2} which implies $\text{Re}(\alpha\beta^*\gamma^*)=0$. With this, we can keep the definition of the orthonormal states in Eq.~\ref{eq:basis_c1}, and use the same steps in the calculation of the trace in Eqs.~\ref{eq:trace_c1} and part of Eq.~\ref{eq:pre_condition_c1} up to $\braket{\Omega_1|\Omega_2} = \pm 1$. In this case the condition is slightly different:
\begin{equation}\label{eq:unitar_c2}
	\braket{\Omega_1|\Omega_2} = \pm 1 \leftrightarrow \bra{\Psi}\frac{\alpha^* I + \beta^* \gamma^* P_B}{\sqrt{1-|\beta|^2|\delta|^2}}\mathcal{U}_1^\dagger\mathcal{U}_2P\ket{\Psi}.
\end{equation}
Again, this must hold for arbitrary $\ket{\Psi}$, so
\begin{equation}\label{eq:condition_c2}
	\bra{\Psi}\frac{\alpha^* I + \beta^* \gamma^* P_B}{\sqrt{1-|\beta|^2| \delta|^2}} \mathcal{U}_1^\dagger\mathcal{U}_2P\ket{\Psi} = I \leftrightarrow \mathcal{U}_2 = \mathcal{U}_1 \frac{\alpha I + \beta \gamma P_B}{\sqrt{1-|\beta|^2| \delta|^2}}.
\end{equation}
One can double check this does indeed disentangle the original state, but we will omit this step. It is however useful to rewrite the unitary in the final form:
\begin{equation}\label{eq:f_unitary_c2}
	\mathcal{U}_f = \mathcal{U}_1 \otimes I \left( I\otimes \ket{\omega}\bra{\phi_n} + \frac{\alpha^* P_B + \beta^* \gamma^* I}{\sqrt{1-|\beta|^2| \delta|^2}} \otimes \ket{\hat{\omega}}\bra{\hat{\phi}_n} \right).
\end{equation}
It is now a bit more involved to see that this cannot be a Clifford. We'll use the trick in Eq.~\ref{eq:last_step_c1} twice. First with the state $\ket{s} = \bigotimes_{i=1}^{n-1}\ket{s_i}$ such that $P_i\ket{s_i}=\ket{s_i}$ and $\ket{s_n}=\ket{0}$. Remembering Eq.~\ref{eq:state_c2} and using that $(\alpha^* + \beta^* \gamma^*)/(\sqrt{1-|\beta|^2| \delta|^2})$ has norm $1$ (see Eq.~\ref{eq:norm_c2}) so we can write it as $e^{i\varphi}$, the action of $\mathcal{U}_f$ on this state is:
\begin{equation}\label{eq:ex1_c2}
	\mathcal{U}_f\bigotimes_{i=1}^{n}\ket{s} = \mu^* \mathcal{U}_1\ket{s} \otimes \ket{\omega} + \nu e^{i\varphi} \mathcal{U}_1 \ket{s} \otimes \ket{\bar{\omega}} = \mathcal{U}_1\ket{s} \otimes (\mu^* \ket{\omega} + \nu e^{i\varphi} \ket{\bar{\omega}} ).
\end{equation}
If this is not a stabilizer state, the proof is done. If it is, then we can assume without loss of generality that the state in qubit $n$ is the $\ket{0}$ state specifically (because changing from one stabilizer state to another is a Clifford transformation that does not affect the reasoning). That is,
\begin{equation}\label{eq:ex1_s0_c2}
	(\mu^* \ket{\omega} + \nu e^{i\varphi} \ket{\bar{\omega}}) = \ket{0}.
\end{equation}
This implies 
\begin{equation}\label{eq:ex1_s1_c2}
	(\nu^* \ket{\omega} - \mu e^{i\varphi} \ket{\bar{\omega}}) = \ket{1},
\end{equation}
and therefore
\begin{equation}\label{eq:ex1_state_c2}
	\begin{split}
		\ket{\omega} &=	\mu \ket{0} + \nu \ket{1} = \ket{\phi_n} \\
		\ket{\bar{\omega}} &= e^{-i\varphi} (\nu^* \ket{0} - \mu^* \ket{1}) = e^{-i\varphi} \ket{\bar{\phi}_n}
	\end{split}\: .
\end{equation}
Now we use instead the stabilizer state that we saw in the simplified proof, $\ket{s} = \bigotimes_{i=1}^{n-1}\ket{s_i}$ such that $P_i\ket{s_i}=\ket{\bar{s}_i}$ (or the same state if $P_i=I$) and $\ket{s_n}=\ket{0}$. As beforem, it is also true that $\braket{s|\bar{s}}=0$. Applying $\mathcal{U}_f$ to this state gives
\begin{equation}\label{eq:ex2_state_c2}
	\mathcal{U}_f\bigotimes_{i=1}^{n}\ket{s_i} = \mu^* \mathcal{U}_1\ket{s} \otimes \ket{\phi_n} - \nu \frac{\alpha e^{-i\varphi}}{\sqrt{1-|\beta|^2|\delta|^2}} \mathcal{U}_1 \ket{\bar{s}} \otimes \ket{\bar{\phi}_n} - \nu \frac{\beta \gamma e^{-i\varphi}}{\sqrt{1-|\beta|^2| \delta|^2}} \ket{s} \otimes \ket{\bar{\phi}_n}.
\end{equation}
Checking the purity of the reduced density matrix $\rho_n$ on qubit $n$ for this state $\rho_L$ is a longer computation, but it can be done nonetheless. Substituting $\ket{r}=U_1\ket{s}$, the state $\rho_L$ is
\begin{equation}\label{eq:final_c2}
	\begin{split}
		\rho_L&= \mathcal{U}_f \ket{s}\otimes\ket{0}\bra{s}\otimes\bra{0}\mathcal{U}^\dagger_f
		= |\mu|^2 \ket{r\phi_n}\bra{r \phi_n} + \frac{|\nu|^2}{\sqrt{1-|\beta|^2|\delta|^2}} |\alpha|^2\ket{\bar{r}\bar{\phi}_n}\bra{\bar{r}\bar{\phi}_n} - \\ 
		&- \left( \mu^* \nu^* \alpha^* \frac{e^{i\varphi}}{\sqrt{1-|\beta|^2|\delta|^2}}\ket{r\phi_n}\bra{\bar{r}\bar{\phi}_n} + \mu^* \nu^* \beta^* \gamma^* \frac{e^{i\varphi}}{\sqrt{1-|\beta|^2|\delta|^2}}\ket{r\phi_n}\bra{r\bar{\phi}_n} + \right. \\
		&+ \left. \mu \nu \alpha \frac{e^{-i\varphi}}{\sqrt{1-|\beta|^2|\delta|^2}}\ket{\bar{r}\bar{\phi}_n}\bra{r\phi_n} + \mu \nu \beta \gamma \frac{e^{-i\varphi}}{\sqrt{1-|\beta|^2|\delta|^2}}\ket{r\bar{\phi}_n}\bra{r\phi_n} \right) ,
	\end{split}
\end{equation}
so the reduced $\rho_n$, using the basis $\{ \ket{r},\ket{\bar{r}} \}$ for the trace, is:
\begin{equation}
	\begin{split}
	\rho_n = \text{Tr}_B\left[ \rho_L \right] &= \bra{r}\rho_L\ket{r} + \bra{\bar{r}}\rho_L\ket{\bar{r}} = \begin{pmatrix}
		|\mu|^2 & \mu^*\nu^*\beta^*\gamma^*\frac{e^{i\varphi}}{\sqrt{1-|\beta|^2|\delta|^2}} \\
		\mu\nu\beta\gamma\frac{e^{-i\varphi}}{\sqrt{1-|\beta|^2|\delta|^2}} & |\nu|^2 
	\end{pmatrix} \rightarrow \\ &\rightarrow \rho_n^2 =
	\begin{pmatrix}
		|\mu|^4 + \frac{|\mu|^2|\nu|^2|\beta|^2|\gamma|^2}{1-|\beta|^2-|\beta|^2|\gamma|^2} & \dots \\
		\dots & |\nu|^4 + \frac{|\mu|^2|\nu|^2|\beta|^2|\gamma|^2}{1-|\beta|^2-|\beta|^2|\gamma|^2}
	\end{pmatrix} \: .
	\end{split}
\end{equation}
Since the purity, calculated with Tr$(\rho_n^2)$, only cares about the diagonal, we omitted some terms in the right equality. Undoing the initial substitution $1-|\delta|^2=|\gamma|^2$, we get
\begin{equation}\label{eq:purity_func}
	P(\rho_n^2) = |\mu|^4 + |\nu|^4 + \frac{2|\mu|^2|\nu|^2|\beta|^2|\gamma|^2}{1-|\beta|^2+|\beta|^2|\gamma|^2}.
\end{equation}
One can simplify this function by substituting
\begin{equation}\label{eq:subs_c2}
	|\mu|^2 = \cos^2(x) \: , \: |\nu|^2 = \sin^2(x) \: , \: |\beta|^2 = \sin^2(y) \: , \: |\gamma|^2 = \cos^2(z) ,
\end{equation} 
which lets us find some clear bounds to this function and under which conditions they are satisfied. First observe that $\sin^2(y)\leq 1$ means that
\begin{equation}\label{eq:up_bound_p1}
	\frac{1-|\beta|^2}{|\beta|^2|\gamma|^2} \geq 0 .
\end{equation}
The rightmost term in Eq.~\ref{eq:purity_func}, then:
\begin{equation}\label{eq:up_bound_p2}
	\frac{2|\mu|^2|\nu|^2|\beta|^2|\gamma|^2}{1-|\beta|^2+|\beta|^2|\gamma|^2} = \frac{2|\mu|^2|\nu|^2}{1+\frac{1-|\beta|^2}{|\beta|^2|\gamma|^2}} \leq 2|\mu|^2|\nu|^2,
\end{equation}
where we used Eq.~\ref{eq:up_bound_p1} in the last inequality, which is only satisfied if $\mu$ or $\nu$ are $0$ or if $|\beta|^2=1$. In the other cases, we have
\begin{equation}\label{eq:up_bound}
	P(\rho_n^2) < |\mu|^4 + |\nu|^4 + 2|\mu|^2|\nu|^2| = (|\mu|^2 + |\nu|^2)^2 = (\cos^2(x) + \sin^2(x))^2 = 1.
\end{equation}
On the other hand, $|\beta|^2|\gamma|^2\in[0,1]$ means this same rightmost term we upper bounded is always greater than $0$, so
\begin{equation}\label{eq:low_bound}
	P(\rho_n^2) \geq |\mu|^4 + |\nu|^4 = \cos^4(x) + \sin^4(x) \geq \frac{1}{2} .
\end{equation}
One can check that this lower bound is only reached if $x=\pi/4$. As we saw in the simplified proof, if the state is a stabilizer state the reduced matrix of a single qubit must be either pure or maximally mixed, which means the purity can only be $1$ or $1/2$ respectively. Considering that the bounds we found are only reached when $\phi_n$ is a stabilizer state ($x=\pi/4$,$\mu=0$ or $\nu=0$) or when the initial rotation is a Clifford rotation in the first place ($|\beta|^2=1$), the resulting state cannot be a stabilizer in any other scenario (other than those observed in the theorem), and thus the unitary cannot be a Clifford gate, completing the proof. $\qed$ 

\end{document}